# MODELING ORBITAL GAMMA-RAY SPECTROSCOPY EXPERIMENTS AT CARBONACEOUS ASTEROIDS.

LUCY F. LIM[1], RICHARD D. STARR[1,2], LARRY G. EVANS[1,3], ANN M. PARSONS[1], MICHAEL E. ZOLENSKY[4], WILLIAM V. BOYNTON[5]

[1]NASA Goddard Space Flight Center
[2]Catholic University of America
[3]Computer Sciences Corporation
[4]ARES, NASA Johnson Space Center, Houston, TX 77058
[5]University of Arizona

## ABSTRACT

To evaluate the feasibility of measuring differences in bulk composition among carbonaceous meteorite parent bodies from an asteroid or comet orbiter, we present the results of a performance simulation of an orbital gamma-ray spectroscopy ("GRS") experiment in a Dawn-like orbit around spherical model asteroids with a range of carbonaceous compositions. The orbital altitude was held equal to the asteroid radius for 4.5 months. Both the asteroid $\gamma$-ray spectrum and the spacecraft background flux were calculated using the MCNPX Monte-Carlo code.

GRS is sensitive to depths below the optical surface (to ≈20–50 cm depth depending on material density). This technique can therefore measure underlying compositions beneath a sulfur-depleted (*e.g.*, *Nittler et al. 2001*) or desiccated surface layer.

We find that $3\sigma$ uncertainties of under 1 wt% are achievable for H, C, O, Si, S, Fe, and Cl for five carbonaceous meteorite compositions using the heritage Mars Odyssey GRS design in a spacecraft-deck-mounted configuration at the Odyssey end-of-mission energy resolution, FWHM = 5.7 keV at 1332 keV. The calculated compositional uncertainties are smaller than the compositional differences between carbonaceous chondrite subclasses.

## 1. CARBONACEOUS METEORITES AND THEIR PARENT BODIES

Carbonaceous meteorites differ substantially from one another in degree of hydration and in their bulk abundances of carbon, sulfur, and various other elements. Levels of hydration in carbonaceous chondrites range from >17 wt% in the CI chondrites (*Wiik* 1956; *Jarosewich* 1990) to essentially zero in the CK and reduced CV subclasses. Bulk carbon (Fig. 1) can range from <0.5% in the CK chondrites to > 4% in Tagish Lake (*Pearson et al.* 2006; *Grady et al.* 2002) and bulk sulfur from <1% in CKs to >5 wt% in CI chondrites (*Jarosewich* 1990).

We evaluated the feasibility of measuring these differences in bulk composition from an asteroid or comet orbiter by conducting a performance simulation of an orbital gamma-ray spectroscopy ("GRS") experiment in a Dawn-like orbit around five spherical model asteroids with carbonaceous meteoritic compositions. As in the Dawn Vesta encounter as planned (e.g. *Russell et al.* 2004) the orbital altitude was held equal to the asteroid radius for 4.5 months. Because GRS is sensitive to depths below the optical surface (to ≈20–50 cm depth depending on material density), it can provide bulk compositions of material hidden from other spectral techniques beneath a sulfur-depleted (*e.g.*, *Nittler et al. 2001*) or desiccated surface layer.

lucy.f.lim@nasa.gov

## 2. METHODS

### 2.1. *Monte-Carlo Model*

We applied the well-established Monte-Carlo radiation transport code MCNPX (*Pelowitz* 2005; *Waters et al.* 2007) to simulate natural galactic cosmic-ray ("GCR")-induced gamma-ray production in the surfaces of five model asteroids with compositions based on each of four types of carbonaceous chondrites and on a ureilite (a carbonaceous achondrite). Key elemental abundances are given in Table 1 and full compositions in Supplemental Table S1.

We then modeled a spacecraft background based on a Dawn-like spacecraft model, also using MCNPX, as described below (§ 2.3). Finally, MCNPX was used to model the interaction of the combined asteroid and spacecraft spectra with the heritage Mars Odyssey GRS instrument (*Boynton et al.* (2004); § 2.5), and the resulting spectra analyzed (§ 2.6) to calculate sensitivity to key elements of geochemical interest (Table 2).

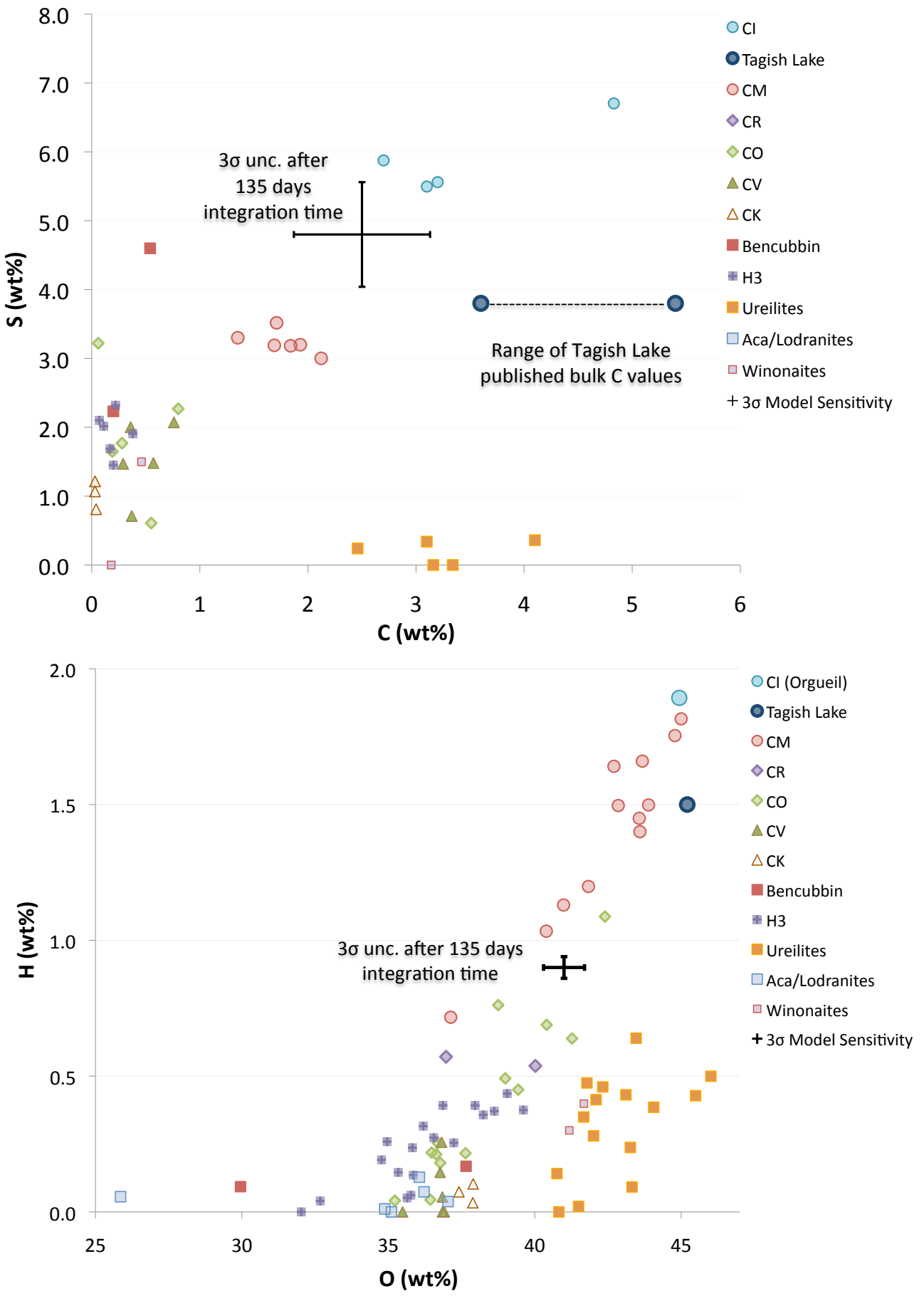

**Figure 1.** *Top:* Orbital gamma-ray spectroscopy experiments can measure bulk sulfur and carbon at the ≈1-5 wt% levels found in carbonaceous meteorites. *Bottom:* GRS is sensitive to bulk hydrogen and oxygen. Tagish Lake compositional data from *Brown et al.* (2000) and *Grady et al.* (2002); other meteorites as compiled by *Nittler et al.* (2004), including data from *Wiik* (1956); *Jarosewich* (1990); *Yanai and Kojima* (1995). Sensitivities reflect the end-of-mission energy resolution of the Odyssey GRS, full-width-at-half-maximum (FWHM) = 5.7 keV at 1332 keV. See Table 2.

**Table 1.** Compositional Models Based on Carbonaceous Meteorites

| Z | Element | Abundance (wt%) | | | | |
|---|---|---|---|---|---|---|
| | | CI | Tagish Lake | CM | CO | Novo Urei |
| 1 | H | 2.02 | 1.5 | 1.4 | 0.07 | 0 |
| 6 | C | 3.45 | 3.6[a] | 2.2 | 0.45 | 2.23 |
| 8 | O | 46.1 | 45.2 | 43.2 | 37.0 | 38.7 |

*Table 1 continued*

Table 1 *(continued)*

| Z | Element | Abundance (wt%) | | | | |
|---|---|---|---|---|---|---|
| | | CI | Tagish Lake | CM | CO | Novo Urei |
| 14 | Si | 10.64 | 11.4 | 12.7 | 15.9 | 18.57 |
| 16 | S | 5.41 | 3.8 | 2.7 | 2.0 | 0.58 |
| 26 | Fe | 18.2 | 19.3 | 21.3 | 24.8 | 15.64 |

[a]Carbon wt% from Table 2 of *Brown et al.* (2000). Higher bulk carbon abundances for this meteorite are reported in the text of *Brown et al.* (2000) and in other papers (*e.g.* Grady et al. 2002, Pearson et al., 2006)

NOTE—Meteorite models were based on published analyses or averages as follows: CI, Tagish Lake, CM— *Brown et al.* (2000); CO—Subclass average from *Wasson and Kallemeyn* (1988); Novo Urei— Major elements from *Wiik* (1972); REE and trace elements from *Goodrich et al.* (1991); *Spitz and Boynton* (1991).

### 2.2. *Cosmic-ray flux source term*

The integral 4$\pi$-steradian flux of galactic cosmic ray ("GCR") protons varies from 1.5 to 5.0 particles/cm$^2$/s over the course of the 11-year solar cycle (*McKinney et al.* 2006). To be conservative, we normalized our model GCR flux to high solar potential (low GCR flux), comparable to that seen by the Mars Odyssey GRS during its first three years in Mars orbit. Missions that orbit instead during solar minimum may thus see gamma-ray fluxes up to three times higher than those described here.

We normalized the GCR flux on the planet's surface as in *McKinney et al.* (2006), dividing the isotropic GCR flux by a factor of 4, which provides the proper weighting for a surface source flux in MCNPX. For a solar potential of 1000 (solar maximum), *McKinney et al.* have GCR proton and alpha 4$\pi$ fluxes of 1.5 and 0.1 particles/cm$^2$/s, respectively. Following *McKinney et al.* and others (*e.g. Masarik and Reedy 1996)* we accounted for the alpha contribution by multiplying the alpha flux by a factor of 3.8, giving us an effective isotropic GCR proton flux of 1.5 + 0.38 = 1.88 particles/cm$^2$/s, which corresponds to 0.47 particles/cm$^2$/s on the asteroid surface. (We note, however, that the 25% contribution of the alpha particle component remains minor in comparison with the ≈300% variation caused by the solar cycle.)

**Table 2**. Summary Orbital GRS Single-Line Sensitivities after 4.5 Months (135d) Integration Time

| $\gamma$-Ray | Accumulated $\gamma$-Ray Count Unc. ($3\sigma$) after 135 days Int. Time | | Element Wt% in Model[a] | Idealized Compositional Uncertainty (wt%; $3\sigma$) | |
|---|---|---|---|---|---|
| | At Energy Resolution of: | | | At Energy Resolution of: | |
| | 4.1 keV | 5.7 keV | | 4.1 keV | 5.7 keV |
| Neutron Capture $\gamma$-Rays (n,$\gamma$) | | | | | |
| H 2223 keV[b] | | | | | |
| CI | ± 0.80% | ± 1.17% | 2.02 wt% | ± 0.016 wt% | ± 0.024 wt% |
| Tagish Lake | ± 1.01% | ± 1.15% | 1.50 wt% | ± 0.015 wt% | ± 0.017 wt% |
| CM | ± 1.12% | ± 1.35% | 1.40 wt% | ± 0.016 wt% | ± 0.019 wt% |
| CO | ± 30.1% | ± 58 % | 0.07 wt% | ± **0.021 wt%** | ± **0.040 wt%** |
| Novo Urei | ··· | ··· | 0 wt% | ··· | ··· |
| Si 4934 keV | | | | | |
| CI | ± 4.6% | ± 5.8% | 10.64 wt% | ± 0.49 wt% | ± 0.62 wt% |
| Tagish Lake | ± 4.3% | ± 5.3% | 11.40 wt% | ± 0.49 wt% | ± 0.61 wt% |
| CM | ± 4.3% | ± 5.1% | 12.70 wt% | ± 0.54 wt% | ± 0.65 wt% |
| CO | ± 5.4% | ± 6.6% | 15.90 wt% | ± **0.86 wt%** | ± **1.04 wt%** |
| Novo Urei | ± 4.3% | ± 5.2% | 18.57 wt% | ± 0.80 wt% | ± 0.97 wt% |
| S 5420 keV | | | | | |
| CI | ± 4.3% | ± 5.3% | 5.41 wt% | ± 0.23 wt% | ± 0.29 wt% |
| Tagish Lake | ± 6.1% | ± 8.3% | 3.80 wt% | ± 0.23 wt% | ± 0.31 wt% |
| CM | ± 8.9% | ± 13.5% | 2.70 wt% | ± 0.24 wt% | ± 0.36 wt% |
| CO | ± 23 % | ± 38 % | 2.00 wt% | ± **0.46 wt%** | ± **0.76 wt%** |
| Novo Urei[c] | ± 99.6% | ··· | 0.58 wt% | ± 0.58 wt% | ··· |
| Cl 6111 keV | | | | | |
| CI | ± 7.8% | ± 9.4% | 0.070 wt% | ± 0.0054 wt% | ± 0.0066 wt% |
| Tagish Lake | ± 9.4% | ± 11.5% | 0.056 wt% | ± 0.0053 wt% | ± 0.0065 wt% |
| CM | ± 11.9% | ± 14.2% | 0.043 wt% | ± 0.0051 wt% | ± 0.0061 wt% |
| CO | ± 34 % | ± 41 % | 0.024 wt% | ± **0.0083 wt%** | ± **0.0098 wt%** |
| Novo Urei | ± 53 % | ± 84 % | 0.008 wt% | ± 0.0041 wt% | ± 0.0064 wt% |
| Fe 7632 keV | | | | | |
| CI | ± 1.3% | ± 1.8% | 18.20 wt% | ± 0.23 wt% | ± 0.33 wt% |
| Tagish Lake | ± 1.2% | ± 1.4% | 19.30 wt% | ± 0.24 wt% | ± 0.27 wt% |
| CM | ± 1.2% | ± 1.6% | 21.30 wt% | ± 0.26 wt% | ± 0.34 wt% |
| CO | ± 1.6% | ± 1.8% | 24.80 wt% | ± **0.41 wt%** | ± **0.46 wt%** |
| Novo Urei | ± 1.9% | ± 2.1% | 15.64 wt% | ± 0.30 wt% | ± 0.34 wt% |

*Table 2 continued*

Table 2 *(continued)*

| γ-Ray | Accumulated γ-Ray Count Unc. (3σ) after 135 days Int. Time | | Element Wt% in Model[a] | Idealized Compositional Uncertainty (wt%; 3σ) | |
|---|---|---|---|---|---|
| | At Energy Resolution of: | | | At Energy Resolution of: | |
| | 4.1 keV | 5.7 keV | | 4.1 keV | 5.7 keV |
| Inelastic Scatter γ-Rays (n,n′γ) | | | | | |
| Si 1779 keV[d] | | | | | |
| CI | ± 2.6% | ± 2.9% | 10.64 wt% | ± **0.28 wt%** | ± **0.31 wt%** |
| Tagish Lake | ± 2.3% | ± 2.6% | 11.40 wt% | ± 0.27 wt% | ± 0.30 wt% |
| CM | ± 2.1% | ± 2.4% | 12.70 wt% | ± 0.27 wt% | ± 0.30 wt% |
| CO | ± 1.2% | ± 1.3% | 15.90 wt% | ± 0.19 wt% | ± 0.21 wt% |
| Novo Urei | ± 1.1% | ± 1.2% | 18.57 wt% | ± 0.20 wt% | ± 0.22 wt% |
| C 4438 keV[e] | | | | | |
| CI | ± 14.3% | ± 18.2% | 3.45 wt% | ± 0.49 wt% | ± **0.63 wt%** |
| Tagish Lake | ± 14.2% | ± 16.0% | 3.60 wt% | ± **0.51 wt%** | ± 0.58 wt% |
| CM | ± 19.2% | ± 19.8% | 2.20 wt% | ± 0.42 wt% | ± 0.44 wt% |
| CO | ± 33 % | ± 49 % | 0.45 wt% | ± 0.15 wt% | ± 0.22 wt% |
| Novo Urei | ± 20 % | ± 22 % | 2.23 wt% | ± 0.45 wt% | ± 0.48 wt% |
| O 6129 keV | | | | | |
| CI | ± 1.80% | ± 1.95% | 46.1 wt% | ± **0.83 wt%** | ± **0.90 wt%** |
| Tagish Lake | ± 1.75% | ± 1.91% | 45.2 wt% | ± 0.79 wt% | ± 0.86 wt% |
| CM | ± 1.82% | ± 1.98% | 43.2 wt% | ± 0.79 wt% | ± 0.86 wt% |
| CO | ± 1.78% | ± 1.92% | 37.0 wt% | ± 0.66 wt% | ± 0.71 wt% |
| Novo Urei | ± 1.65% | ± 1.80% | 38.7 wt% | ± 0.64 wt% | ± 0.70 wt% |
| Radiogenic γ-Rays | | | | | |
| K 1461 keV | | | | | |
| CI | ± 1.15% | ± 1.28% | 0.056 wt% | ±0.00064 wt% | ±0.00072 wt% |
| Tagish Lake | ± 1.04% | ± 1.15% | 0.065 wt% | ±**0.00068 wt%** | ±**0.00075 wt%** |
| CM | ± 1.35% | ± 1.50% | 0.037 wt% | ±0.00050 wt% | ±0.00056 wt% |
| CO | ± 1.15% | ± 1.24% | 0.040 wt% | ±0.00046 wt% | ±0.00050 wt% |
| Novo Urei | ± 1.28% | ± 1.39% | 0.030 wt% | ±0.00038 wt% | ±0.00042 wt% |

*Table 2 continued*

Table 2 *(continued)*

| $\gamma$-Ray | Accumulated $\gamma$-Ray Count Unc. ($3\sigma$) after 135 days Int. Time | | Element Wt% in Model[a] | Idealized Compositional Uncertainty (wt%; $3\sigma$) | |
|---|---|---|---|---|---|
| | At Energy Resolution of: | | | At Energy Resolution of: | |
| | 4.1 keV | 5.7 keV | | 4.1 keV | 5.7 keV |

NOTE—"Idealized Compositional Uncertainty" represents the best possible uncertainty in derived elemental abundance from a single $\gamma$-ray line based solely on counting statistics. It is the product of the model wt% with the line uncertainties and thus assumes a perfect knowledge of a linear relationship between the $\gamma$-ray's count rate and the abundance of its source element. This limiting uncertainty can vary with composition. For each $\gamma$-ray, the composition for which this element's uncertainty is the highest has been marked in bold.

[a] See Table 1 for sources of meteorite compositional data.

[b] In addition to total line strength uncertainties, H 2223 keV uncertainties incorporate (in a root-sum-squared sense) uncertainties in the subtracted line background from hydrazine fuel in the spacecraft model as determined by analysis of a spacecraft-only spectrum. This background contributes approximately 0.03 counts per minute to the total line strength.

[c] Marginal detection at 4.1 keV resolution; no detection at 5.7 keV resolution.

[d] Si 1779 keV uncertainties include uncertainties in the subtracted (≈0.25 counts per minute) line background from Al in the spacecraft model.

[e] C 4438 keV values are statistics in the difference between the Doppler-broadened 4438 keV fitted peak and the predicted counts in this line due to (1) spacecraft background emission from carbon and (2) asteroidal oxygen emission in the 4438 keV line, which is determined from the O 6129 keV line and thus incorporates uncertainties propagated from the 6129 keV line. See Table 6.

### 2.3. *The Spacecraft Model*

The model spacecraft was a Dawn-like solar-electric propulsion design. The spacecraft mass is dominated by aluminum (615 kg). Xenon tanks for a solar-electric system were included together with smaller hydrazine tanks for chemical attitude control. The stainless fuel tanks added a small amount of iron to the model. The tanks were modeled as full, with 538 kg Xe and 50 kg of $N_2H_4$. A 19 kg carbon-fiber antenna was also included; this represented the only onboard carbon in the model.

There was no oxygen or silicon in the spacecraft model. Additionally, although fairly massive, the Dawn solar panels were far from the GRS and preliminary calculations indicated that their contribution to the GRS background would be minimal, whereas including them would have increased the Monte-Carlo running time significantly.

#### 2.3.1. *GCR-induced and neutron-induced spacecraft background*

Background spectra were produced both by direct GCR/spacecraft interactions and by the interaction of the GCR-induced asteroid neutron emission with the spacecraft. The GCR-induced spacecraft background can be measured directly during cruise, but the neutron flux from the asteroid only produces the neutron-induced background during proximity operations.

The asteroid spectrum and both spacecraft background spectra are shown in Figures 2 and 3. Note that at almost all energies the GCR-induced background dominates over the neutron-induced background. This is in contrast to the spacecraft background in planetary mission GRS data sets such as the low-altitude MESSENGER data (*Goldsten et al.* 2007; *Evans et al.* 2012) and in the Mars Odyssey orbital data: the neutron flux received by the spacecraft is proportional to the solid angles occupied by the planets from the point of view of the spacecraft, which were much higher than is practical for an asteroid orbiter.

### 2.4. *Compensating for limitations of the MCNPX Code*

The spectra in this work were initially generated using version 2.6f of MCNPX with the default libraries. As the $\gamma$-ray production cross-sections are not known for their reliability, we carefully checked and corrected the photon output of the code as described below. In addition, the count rates of the $\gamma$-rays of interest in the model spectra were compared to the flight data from the Mars Odyssey GRS experiment (§3.2, Tables 7 and 8).

#### 2.4.1. *Identification and Removal of Spurious Lines*

Although the MCNPX code is reliable in generating neutron fluxes and energy distributions, the $\gamma$-ray spec-tra produced by the code suffer from certain inaccuracies. Spurious lines generated by MCNPX had to be removed from the model spacecraft spectrum —notably lines in both the GCR-induced and neutron-induced spacecraft background that were generated by MCNPX modeling of GCR interactions with aluminum. We verified that these lines were separate from the genuine signal by generating single-element spectra of GCR interactions with various elements including aluminum, oxygen, and carbon. We also verified that these lines were not evident in the cruise data from the Mars Odyssey and MESSENGER missions.

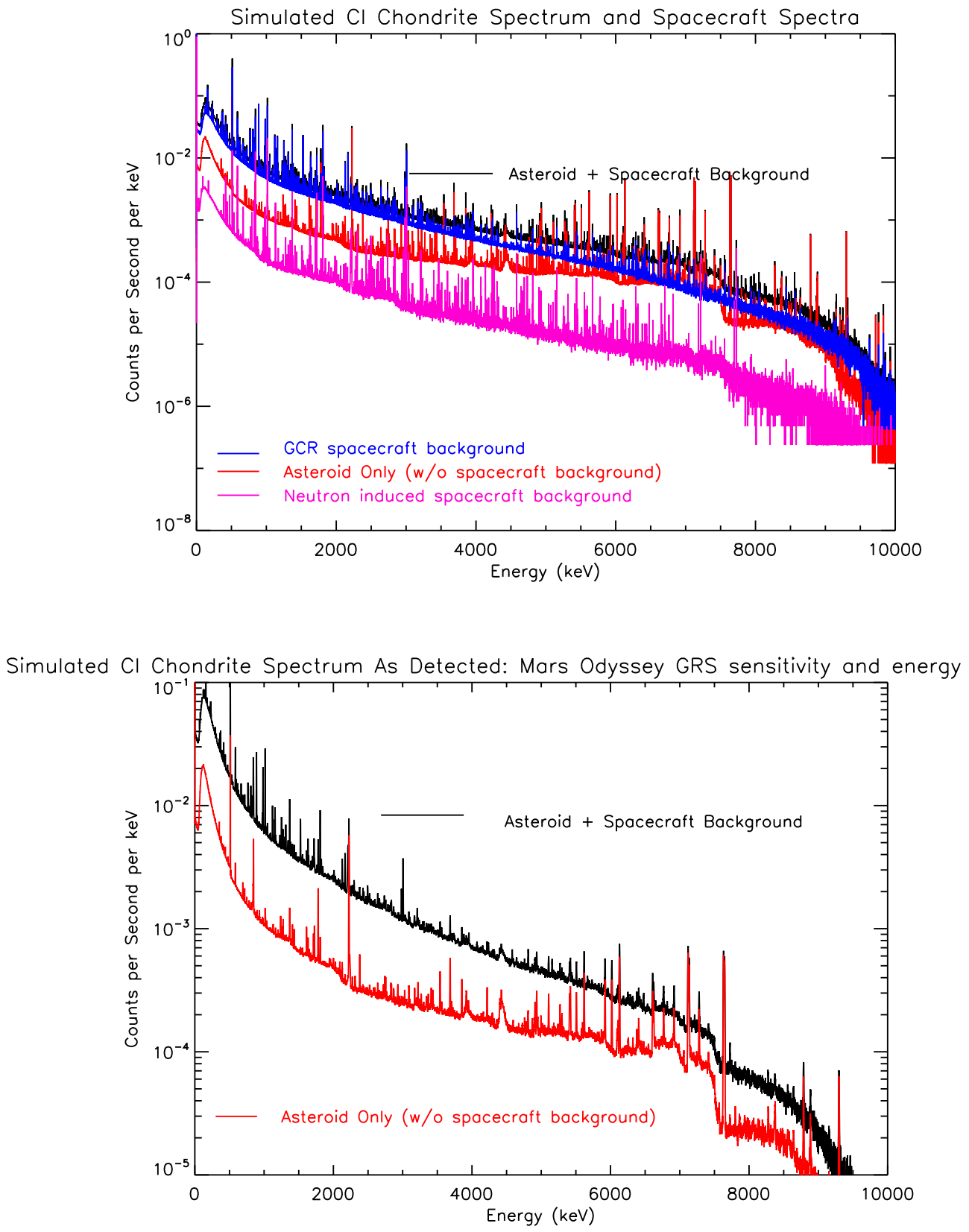

**Figure 2.** Top: CI chondrite gamma-ray spectrum as generated by MCNPX and detected from a Dawn-like low-altitude orbit. Doppler broadening (Table 3) has already been applied to these spectra but the detector resolution has not. Bottom: The same spectra, but as would be detected by the Mars Odyssey detector at its prime mission energy resolution, FWHM = 4.1 keV at 1332 keV (as with all HPGe detectors, $\Delta E \propto \sqrt{E}$). See §2.5.

Note that we did not search the entire MCNPX-generated spectrum for spurious lines. Only the lines in our scientific regions of interest were closely examined. A full list of removed lines is given in Supplemental Table S2.

### 2.4.2. *Identification of Inaccurate Line Energies*

We identified several inaccurate line energies in the MCNPX output. The 2131 keV (n,n′$\gamma$) line of sulfur is inaccurately tallied at 2136 keV. The 1951 keV line of chlorine was useful for the Mars Odyssey GRS analysis but is unfortunately not accurately represented by MCNPX. We based our Cl sensitivity estimates on the stronger 6111 keV neutron capture count rate instead.

### 2.4.3. *Branching Ratios*

The oxygen branching ratio used by MCNPX places too many counts in the 4438 keV line (*i.e.* the 4438 / 6129 keV ratio is too high; see *Peplowski et al.* (2015)). Thus, we are slightly overestimating (Table 6) the uncertainty in carbon based on the oxygen counts in this line.

### 2.4.4. *Application of Doppler Broadening*

Several important spectral lines produced by inelastic scattering of neutrons, such as the 2211 keV Al(n,n′$\gamma$) line, the 2231 keV S(n,n′$\gamma$) line and the 4438 keV $^{12}$C (n,n′$\gamma$) + $^{16}$O(n,n′$\alpha\gamma$)$^{12}$C line, are always Doppler broadened by the nuclear recoil of their production. However, the MCNPX code itself was not used to account for this physical effect— although the neutron scattering was modeled, the entire $\gamma$-ray flux from the reaction was deposited in a single energy channel. Thus, each MCNPX simulation produced a line spectrum superimposed on a continuum (Fig. 2).

In order to accurately simulate a flight spectrum, we applied Doppler broadening to the relevant MCNPX-generated lines (Table 3). This was done prior to modeling the $\gamma$-ray/detector interaction and applying the detector energy resolution function to the full spectrum (§ 2.5).

**Table 3.** Doppler-Broadened $\gamma$-Ray Lines in Regions of Interest

| Line Energy (keV) | Doppler-Broadened FWHM (keV) | Reaction |
|---|---|---|
| 2211 | 14.5 | $^{27}$Al(n,n′$\gamma$)$^{27}$Al |
| 2236 | 22.9 | $^{32}$S(n,n′$\gamma$)$^{32}$S † |
| 4438 | 81.3 | $^{12}$C(n,n′$\gamma$)$^{12}$C + $^{16}$O(n,n′$\alpha\gamma$)$^{12}$C |

†The actual energy of this line is 2231 keV but it is produced by MCNPX in the model spectrum at 2236 keV.

## 2.5. *Detector Efficiency and Resolution*

Finally, the same Monte-Carlo code was used to model the interaction between the summed source and background spectra and the gamma-ray detector, which we based on the most sensitive heritage flight GRS experiment, the Mars Odyssey Gamma-Ray Spectrometer (*Boynton et al.* 2004). The large detector size and high energy resolution (0.3% at 1332 keV during its prime mission) of this instrument give it the best sensitivity of any heritage GRS instrument to carbon, sulfur, and other minor elements in carbonaceous asteroids (Fig. 12). However, we modeled a deck-mounted instrument configuration (similar to that of the Mercury MESSENGER GRS experiment) in which the spacecraft background (§2.3) is much higher than in the boom-mounted Mars Odyssey flight experiment.

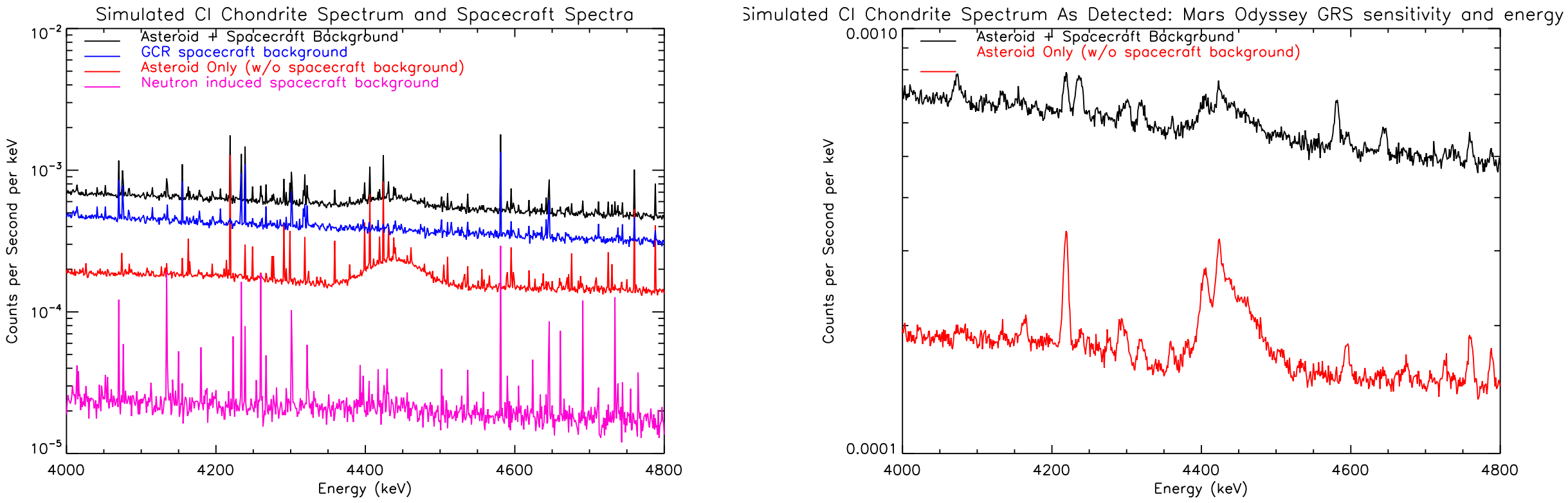

**Figure 3**. Details of spectra shown in Fig. 2 illustrating the Doppler-broadened 4438 keV line. Left: MCNPX-generated spectra after application of Doppler broadening but without detector sensitivity or resolution functions. This corresponds to the energy distribution of photons arriving at the detector. Right: spectra as detected by Mars Odyssey GRS, ΔE (FWHM) = 4.1 keV at 1332 keV.

#### 2.5.1. *Attenuation in the detector housing*

In order to represent attenuation in the detector housing (including the passive cooler), the gamma-rays were simulated as passing through a 2 cm-thick titanium plate prior to entry into the detector. This simplified model of the detector housing was validated by comparison with Mars Odyssey flight data.

#### 2.5.2. *Detector Energy Resolution*

After the above corrections, the last step in preparing the simulated GRS spectrum (Fig. 2) was applying the detector energy resolution function, $\Delta E \propto \sqrt{E}$, as described below.

*Modeling the Effects of Long-Term Radiation Exposure on Detector Resolution* — During the Odyssey mission, the GRS detector was successfully annealed multiple times to repair degradation in energy resolution due to space radiation exposure. The post-anneal energy resolution was slowly degraded by accumulated radiation damage, from 4.1 keV in June 2002 to 5.7 keV over May 2006-Dec. 2007 (FWHM at 1332 keV). We therefore investigated the sensitivity effects of this loss of resolution by analyzing Odyssey GRS data from late in the mission (Table 4).

We generated asteroid and spacecraft background spectra both at the 4.1 keV resolution of the Odyssey prime mission data and at the 5.7 keV resolution of the "Epoch 3" Odyssey data, analyzed both (§ 2.6), and compared the results.

We note that raising the anneal temperature will increase the efficacy of the annealing process. In the Odyssey GRS design, this may be accomplished by adding a heater to the anneal door. Thus, on future missions, it is likely to be feasible to improve the energy resolution over the Odyssey end-of-mission value even after several years of space exposure.

*Energy Resolution of the Kaguya GRS* — The Kaguya lunar mission carried the largest planetary science HPGe GRS experiment flown to date (*Hasebe et al.* 2008) with a 250 $cm^3$ HPGe detector. The pre-flight energy resolution was established at 3.0 keV at 1332 keV (*Hasebe et al.* 2008) comparable to the pre-flight Mars Odyssey and MESSENGER values (Table 4). However, the energy resolution appears to have deteriorated during and after the month-long journey to lunar orbit (2007-09-14 to 2007-10-18; *Kobayashi et al.* (2010)) with the energy resolution at 1461 keV having been measured at ≈6.5 keV in November 2007 and ≈9.5 keV by February 2008 (Figure 4c of *Hasebe et al.* (2010)). Scaling by $\sqrt{(E)}$, these values would correspond to 1332 keV resolutions of 6.2 and 9.1 keV respectively. Later in the mission, the detector was annealed at nominal temperatures of 80–90°C for 48 hours (*Kobayashi et al.* 2013) but the post-anneal resolution for the period 2009-02-10 to 2009-05-29 was limited to 20 keV at 3.7 MeV (*Yamashita et al.* 2012) or 6–7 keV at 1461 keV (*Kobayashi et al.* 2013).

The Mars Odyssey and MESSENGER instruments

**Table 4**. Energy Resolution of Heritage High-Purity Germanium Detectors in Deep Space Over Time

| γ-ray FWHM at a reference energy of 1332 keV (all values post-anneal). Energy resolution in high-purity germanium detectors varies according to: $\Delta E \propto \sqrt{E}$ | | | | | |
|---|---|---|---|---|---|
| | | Mars Odyssey GRS Typical Anneal 73°C | | MESSENGER GRS† Anneal Temp. 85°C | |
| | Time After Launch | Dates | ΔE | Dates | ΔE |
| Pre-Launch | | Before April 2001 | 2.8 keV | < Aug. 2004 | 3.49 keV |
| Cruise | ≈0.3 years | June–July 2001 | 3.8 keV | Nov. 2004 | 3.7 keV* |
| Mars Odyssey Prime Mission | 1.2 years | June 2002 | 4.1 keV | | |
| Mars Odyssey Ext. Mission Epoch 2 | 4–5 years | Apr. 2005–Mar. 2006 | 5.06 keV* | | |
| Mars Odyssey Ext. Mission Epoch 3 | 5–6.5 years | May 2006–Dec. 2007 | 5.79 keV* | | |
| Mercury MESSENGER Prime Mission | 6.6 years | | | Mar. 2011 | 4.8 keV* |

† *Goldsten et al.* (2007); *Evans et al.* (2012)
*Measurement taken at 1369 keV

were also both in flight during this time but did not experience comparable post-anneal losses of energy resolution. The unusual severity of the Kaguya GRS radiation damage has been attributed to cryocooler stoppages (*Hasebe et al.* 2010; *Kobayashi et al.* 2013).

We note that in spite of the unusual radiation damage, the energy resolution of the Kaguya GRS remained significantly better than that of the various scintillator GRS experiments and thus the Kaguya data remain the highest-energy-resolution lunar GRS spectra to date by a wide margin.

### 2.6. *Peak Fitting and Count Rate Extraction*

When the model spectra were complete, peak fluxes were extracted and uncertainties estimated using the same methods that were applied to the Mars Odyssey flight data (*Evans et al.* 2006). (Fig. 6)

Line count rates as retrieved by the peak-fitting procedure are given in Table 5.

**Table 5**. Model Results: Modeled γ-Ray Line Count Rates and Retrieved Count Rates in Counts per Minute (“CPM”)

| γ Line Model | Line CPM | Retrieved Count Rate (CPM) at 4.1 keV resolution[a] | | Retrieved Count Rate (CPM) at 5.7 keV resolution[a] | | % retrieved[b] at 4.1 keV | % retrieved[b] at 5.7 keV |
|---|---|---|---|---|---|---|---|
| Si 1779 keV[c] | | | | | | | |
| CI | 0.726 | 0.736 | ± 0.0069 | 0.723 | ± 0.0085 | 101.3% | 99.6% |
| Tagish Lake | 0.788 | 0.794 | ± 0.0070 | 0.783 | ± 0.0077 | 100.8% | 99.3% |
| CM | 0.846 | 0.854 | ± 0.0071 | 0.839 | ± 0.0078 | 101.0% | 99.2% |
| CO | 1.358 | 1.372 | ± 0.0080 | 1.346 | ± 0.0085 | 101.0% | 99.1% |
| Novo Urei | 1.570 | 1.585 | ± 0.0082 | 1.559 | ± 0.0089 | 101.0% | 99.3% |

*Table 5 continued*

Table 5 *(continued)*

| γ Line Model | Line CPM | Retrieved Count Rate (CPM) at 4.1 keV resolution[a] | | Retrieved Count Rate (CPM) at 5.7 keV resolution[a] | | % retrieved[b] at 4.1 keV | % retrieved[b] at 5.7 keV |
|---|---|---|---|---|---|---|---|
| H 2223 keV | | | | | | | |
| CI | 1.839 | 1.783 | ± 0.0087 | 1.795 | ± 0.0133 | 97.0% | 97.6% |
| Tagish Lake | 1.350 | 1.303 | ± 0.0078 | 1.359 | ± 0.0087 | 96.5% | 100.7% |
| CM | 1.208 | 1.159 | ± 0.0075 | 1.209 | ± 0.0092 | 95.9% | 100.1% |
| CO | 0.099 | 0.069 | ± 0.0058 | 0.099 | ± 0.0072 | 69.6% | 100.3% |
| Novo Urei[c] | 0.067 | 0.029 | ± 0.0052 | 0.080 | ± 0.0064 | 42.9% | 120.2% |
| Si 4423 keV | | | | | | | |
| CI | 0.038 | 0.046 | ± 0.0028 | 0.056 | ± 0.0056 | 120.7% | 146.0% |
| Tagish Lake | 0.039 | 0.042 | ± 0.0040 | 0.047 | ± 0.0056 | 107.6% | 120.1% |
| CM | 0.041 | 0.044 | ± 0.0040 | 0.048 | ± 0.0055 | 106.4% | 116.3% |
| CO | 0.028 | 0.031 | ± 0.0039 | 0.039 | ± 0.0054 | 113.0% | 138.5% |
| Novo Urei | 0.039 | 0.042 | ± 0.0040 | 0.047 | ± 0.0055 | 107.0% | 119.5% |
| Si 4934 keV | | | | | | | |
| CI | 0.088 | 0.100 | ± 0.0040 | 0.088 | ± 0.0036 | 113.1% | 99.9% |
| Tagish Lake | 0.092 | 0.107 | ± 0.0033 | 0.096 | ± 0.0036 | 116.3% | 104.9% |
| CM | 0.096 | 0.109 | ± 0.0033 | 0.100 | ± 0.0036 | 113.7% | 104.3% |
| CO | 0.067 | 0.081 | ± 0.0031 | 0.075 | ± 0.0035 | 120.7% | 111.6% |
| Novo Urei | 0.091 | 0.106 | ± 0.0032 | 0.097 | ± 0.0036 | 116.2% | 106.3% |
| S 5420 keV | | | | | | | |
| CI | 0.105 | 0.096 | ± 0.0029 | 0.099 | ± 0.0037 | 92.2% | 94.7% |
| Tagish Lake | 0.072 | 0.064 | ± 0.0028 | 0.062 | ± 0.0036 | 89.2% | 86.9% |
| CM | 0.048 | 0.042 | ± 0.0027 | 0.037 | ± 0.0035 | 88.2% | 77.7% |
| CO | 0.020 | 0.015 | ± 0.0025 | 0.012 | ± 0.0033 | 73.9% | 60.8% |
| Novo Urei | 0.008 | 0.003 | ± 0.0024 | ... | | 43.7% | ... |
| Cl 6111 keV[3] | | | | | | | |
| CI | 0.042 | 0.043 | ± 0.0024 | 0.044 | ± 0.0029 | 101.7% | 104.1% |
| Tagish Lake | 0.032 | 0.036 | ± 0.0024 | 0.035 | ± 0.0029 | 112.4% | 110.2% |
| CM | 0.024 | 0.028 | ± 0.0024 | 0.028 | ± 0.0028 | 117.7% | 119.0% |
| CO | 0.007 | 0.009 | ± 0.0021 | 0.009 | ± 0.0027 | 127.8% | 136.6% |
| Novo Urei | 0.003 | 0.006 | ± 0.0022 | 0.004 | ± 0.0027 | 188.8% | 147.5% |
| O 6130 keV | | | | | | | |
| CI | 0.266 | 0.262 | ± 0.0033 | 0.269 | ± 0.0037 | 98.4% | 100.8% |
| Tagish Lake | 0.271 | 0.277 | ± 0.0034 | 0.277 | ± 0.0037 | 102.1% | 102.0% |
| CM | 0.258 | 0.262 | ± 0.0034 | 0.262 | ± 0.0037 | 101.4% | 101.4% |
| CO | 0.264 | 0.259 | ± 0.0033 | 0.268 | ± 0.0036 | 97.9% | 101.4% |
| Novo Urei | 0.289 | 0.294 | ± 0.0034 | 0.294 | ± 0.0037 | 101.7% | 101.5% |

*Table 5 continued*

Table 5 *(continued)*

| γ Line Model | Line CPM | Retrieved Count Rate (CPM) at 4.1 keV resolution[a] | | Retrieved Count Rate (CPM) at 5.7 keV resolution[a] | | % retrieved[b] at 4.1 keV | % retrieved[b] at 5.7 keV |
|---|---|---|---|---|---|---|---|
| Fe 7631 keV | | | | | | | |
| CI | 0.359 | 0.359 | ± 0.0033 | 0.360 | ± 0.0046 | 100.0% | 100.3% |
| Tagish Lake | 0.377 | 0.377 | ± 0.0033 | 0.373 | ± 0.0037 | 100.1% | 99.1% |
| CM | 0.392 | 0.393 | ± 0.0034 | 0.389 | ± 0.0044 | 100.1% | 99.0% |
| CO | 0.233 | 0.235 | ± 0.0027 | 0.233 | ± 0.0030 | 101.0% | 100.1% |
| Novo Urei | 0.181 | 0.182 | ± 0.0025 | 0.181 | ± 0.0028 | 100.9% | 100.1% |
| Fe 7645 keV | | | | | | | |
| CI | 0.313 | 0.315 | ± 0.0031 | 0.314 | ± 0.0044 | 100.4% | 100.2% |
| Tagish Lake | 0.328 | 0.328 | ± 0.0031 | 0.327 | ± 0.0035 | 100.1% | 99.8% |
| CM | 0.341 | 0.342 | ± 0.0032 | 0.341 | ± 0.0043 | 100.2% | 100.1% |
| CO | 0.200 | 0.202 | ± 0.0026 | 0.202 | ± 0.0029 | 100.8% | 100.6% |
| Novo Urei | 0.155 | 0.158 | ± 0.0023 | 0.158 | ± 0.0026 | 101.9% | 101.8% |

[a] Energy resolution (FWHM) at the reference energy, 1332 keV; $\Delta E \propto \sqrt{E}$.
[b] Retrieved using peak fitting techniques of *Evans et al.* (2006).

[c] Includes spacecraft background counts from Al.

[d] The Novo Urei model includes no hydrogen. All counts at 2223 keV are attributable to hydrazine in the spacecraft model.

[e] Sum of 6108 and 6109 keV channels in the MCNPX output.

NOTE—All count rates refer to the combined spacecraft+background spectra. Count rate uncertainties are statistical uncertainties after 30 days of accumulation.

## 2.7. *Line Sensitivity and Compositional Sensitivity*

### 2.7.1. *Inelastic-Scatter Lines*

Inelastic scatter line strength is influenced by the total neutron population, which is more or less a function of average atomic number (Z). However, for chondritic materials we find that the inelastic line strengths are essentially proportional to the abundances of their parent elements.

### 2.7.2. *Neutron Capture Lines*

Neutron capture lines are influenced by the population of thermal (low-energy) neutrons, which are much more likely to be captured than higher-energy neutrons. As light elements thermalize neutrons most effectively, the total concentration of light elements influences the strength of all the neutron capture lines. Conversely, certain elements such as Cl and Fe are strong absorbers of thermal neutrons and thus decrease the strength of all capture lines. Hydrogen both thermalizes high-energy neutrons and absorbs thermal neutrons very effectively.

Because the thermal neutron population affects all capture lines equally, the ratio of two capture lines will be proportional to the ratio of their parent elements.

### 2.7.3. *Decay of radiogenic elements*

The natural radioisotopes of K, Th, and U produce $\gamma$-rays independent of the neutron population. The strengths of these lines are directly proportional to the concentrations of their parent nuclides.

# 3. RESULTS

The resulting sensitivity estimates are given above in Table 2. Specific results are discussed below.

## 3.1. *Discussion of Specific Elements and $\gamma$-ray Lines*

### 3.1.1. *Silicon: Inelastic and Capture Lines*

Silicon is very useful for $\gamma$-ray spectral interpretation because planetary $\gamma$-ray spectra generally include strong Si lines produced both by inelastic scatter ($^{28}$Si(n,n′$\gamma$) at 1779 keV) and by neutron capture ($^{28}$Si(n,$\gamma$)$^{29}$Si at 3539 and 4934 keV). Thus, elements for which only inelastic lines can be measured (*e.g.* C) can be referenced to the 1779 keV line, whereas those for which only capture lines can be measured (*e.g.* H) can be normalized to the capture lines of Si.

Figures 4 and 5 show the count rates in these three lines as a function of model silicon abundance.

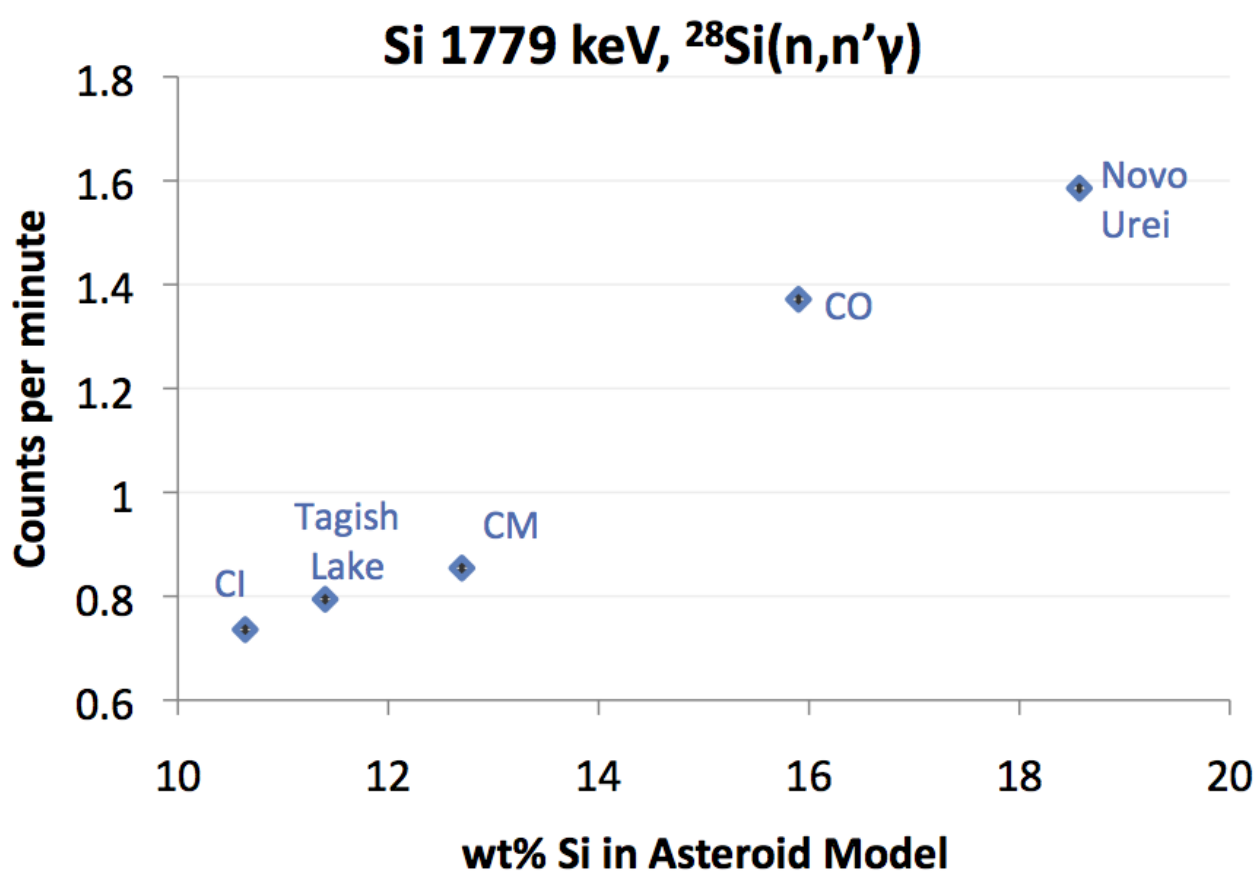

**Figure 4.** **Retrieved count rate in the 1779 keV inelastic line of Si as a function of model asteroid asteroid composition.** Error bars are derived from peak fitting and represent photon statistical errors ($\sigma$) after 30 days of integration. (In this case, the error bars are smaller than the plot symbols (diamonds).)

### 3.1.2. *Hydrogen and the 2223 keV Region*

The 2223 keV $^1$H(n,$\gamma$)$^2$H is the only $\gamma$-ray by which hydrogen can be measured in planetary $\gamma$-ray spectra. Fortunately, the cross-section for this reaction is very high relative to the atomic mass of hydrogen, enabling hydrogen to be measured at low concentrations even in the presence of the overlapping Doppler-broadened spacecraft background from aluminum at 2211 keV and sulfur at 2231 keV. Figure 6 illustrates the position of the 2223 keV line between these two Doppler-broadened peaks.

Figure 7 shows the relationship between 2223 keV count rates and model hydrogen abundances.

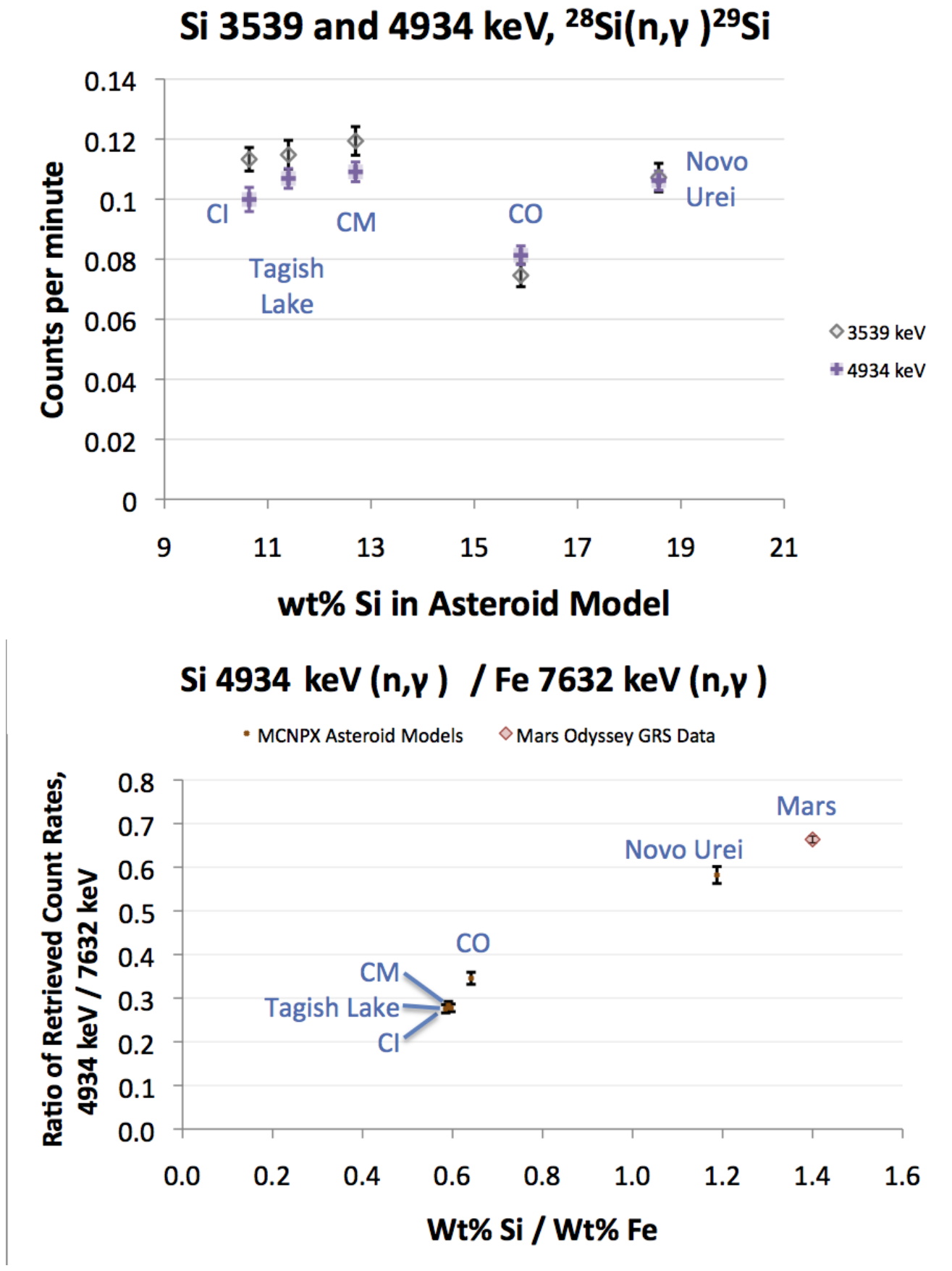

**Figure 5.** **Ratios of neutron capture line strengths correlate with ratios of elemental abundances.** *Top:* Two neutron capture lines of Si. For individual neutron capture lines, the line strengths are not generally linear with the abundance of the emitting element. Capture line strengths depend heavily on how quickly the GCR-produced neutrons are thermalized, which is highly dependent on the light-element abundance in the regolith. However, ratios of one capture line to another are highly correlated with compositional ratios, as can be seen here *(bottom)* in the ratios of neutron capture lines produced by silicon and iron as a function of composition. The Mars Odyssey midlatitude count rate ratio (*Evans et al.* 2006) is also plotted for reference.

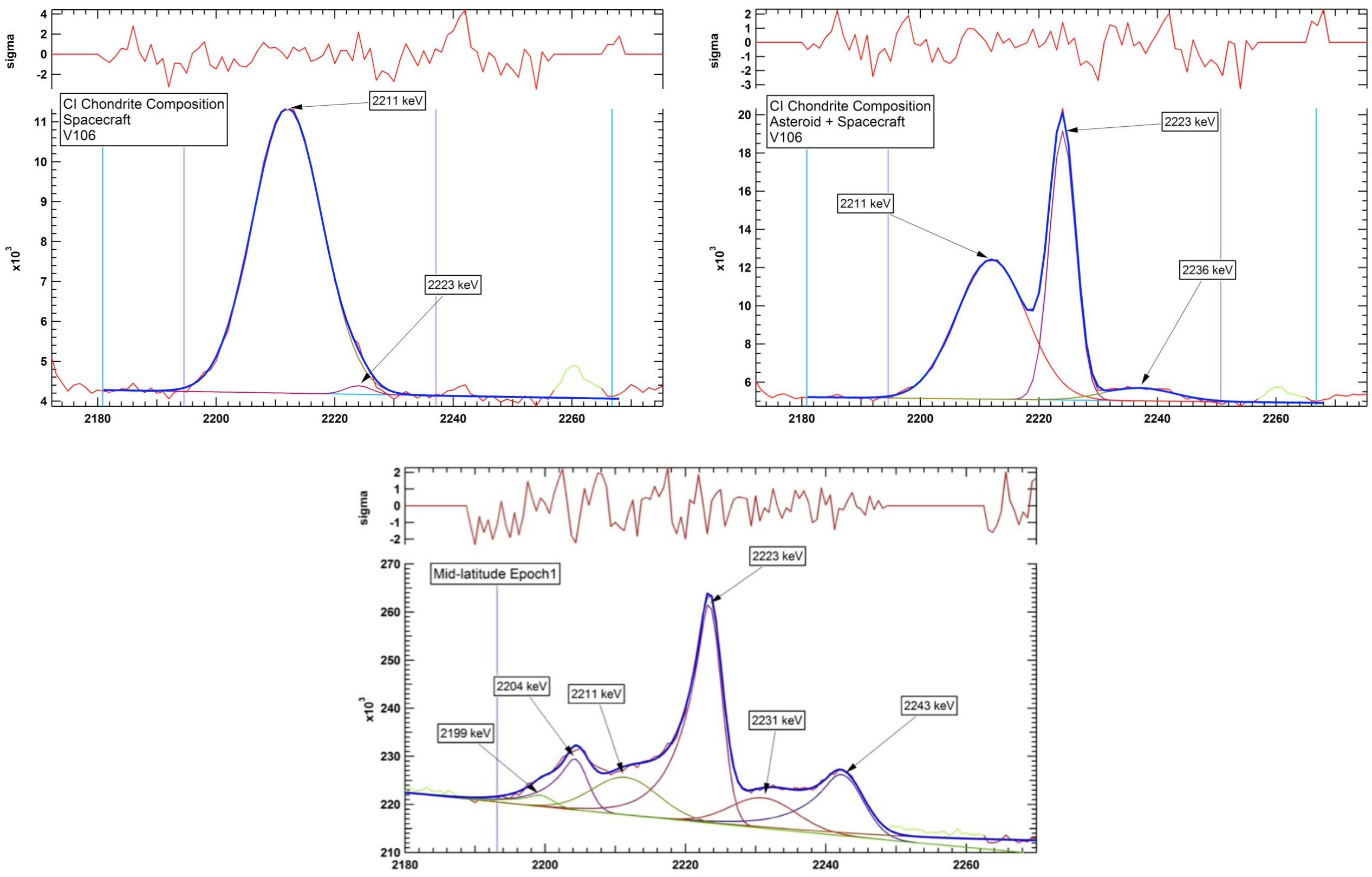

**Figure 6. Fitting of the gamma-ray lines in the complex hydrogen region (2223 keV): model spectra vs. Mars Odyssey data** *Top Left:* Spacecraft-only spectrum. The large Doppler-broadened peak at 2211 keV is aluminum from the spacecraft background. *Top Right:* Fit to the CI chondrite asteroid spectrum with spacecraft background. In spite of the high background, the hydrogen line is sufficiently strong (1.8 cpm) to be measured to well under 1% in flux within 30 days, corresponding to $\ll$0.1 wt% in abundance. Note that the high energy resolution provided by the HPGe detector (0.3%) is required to distinguish these lines. *Bottom:* The same spectral region in the Mars Odyssey mid-latitude data. The count rate in this line was 1.08 counts per minute (*Evans et al.* 2006). As with Mars Odyssey, future missions may increase the hydrogen sensitivity of gamma-ray experiments by mounting the $\gamma$-ray sensor on a boom, thus greatly reducing the spacecraft-originating Al background at 2211 keV.

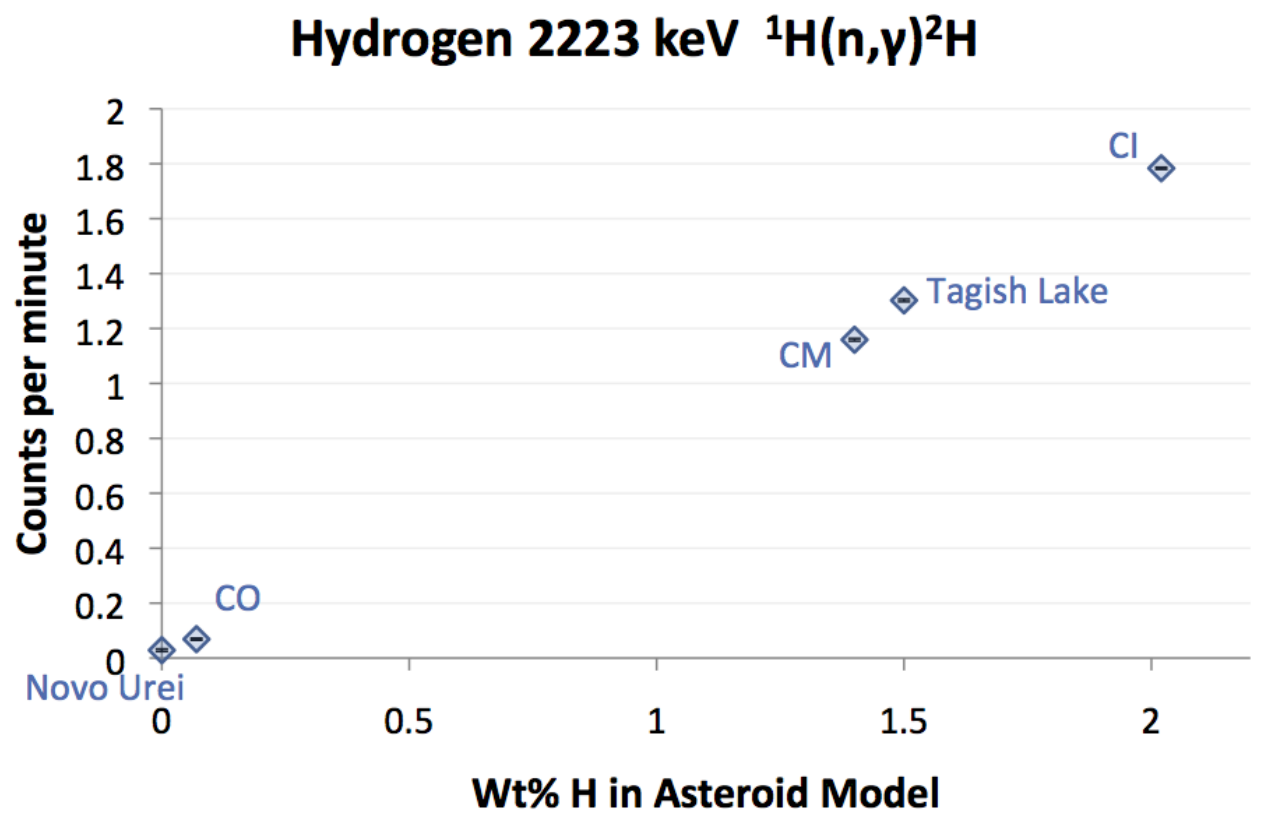

Figure 7. **Retrieved count rate in the 2223 keV line of hydrogen vs. model asteroid composition.** Error bars are derived from peak fitting in Fig. 6 and represent photon statistical errors ($\sigma$) after 30 days of integration. In this case, the error bars are smaller than the plot symbols (diamonds).

#### 3.1.3. *Sulfur*

Sulfur sensitivity was estimated on the basis of the 5420 keV neutron capture line. Figure 8 illustrates count rate in this line as a function of model sulfur abundance.

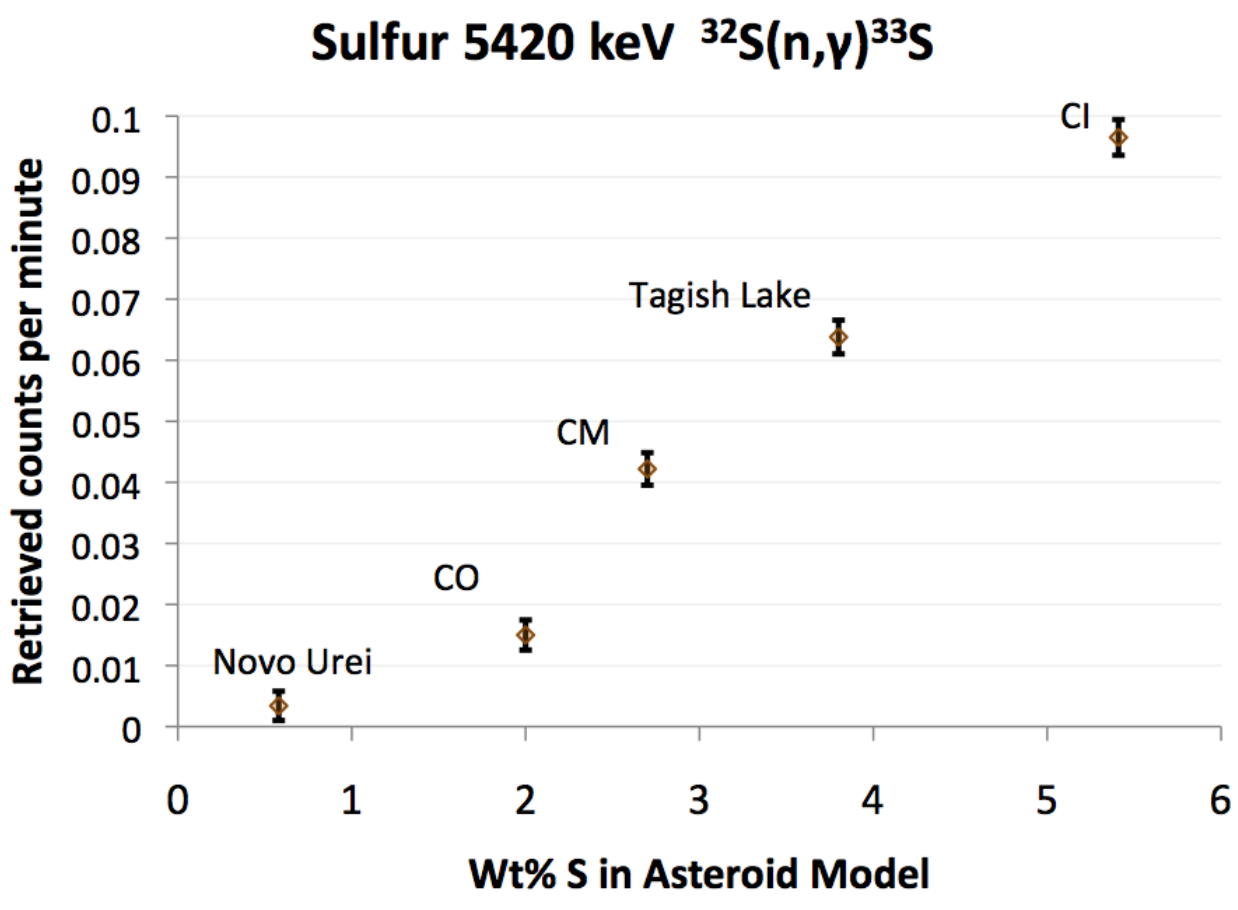

**Figure 8. Retrieved count rate in the 5420 keV line of sulfur vs. model asteroid composition.** Error bars are derived from peak fitting and represent photon statistical errors ($\sigma$) after 30 days of integration.

#### 3.1.4. *Chlorine*

Chlorine is a trace element (<0.1 wt%) in chondrites but thanks to its unusually high neutron capture cross section it can nevertheless easily be measured by an orbital $\gamma$-ray experiment. Figure 9 illustrates count rate and sensitivity in the 6111 keV line as a function of chlorine abundance.

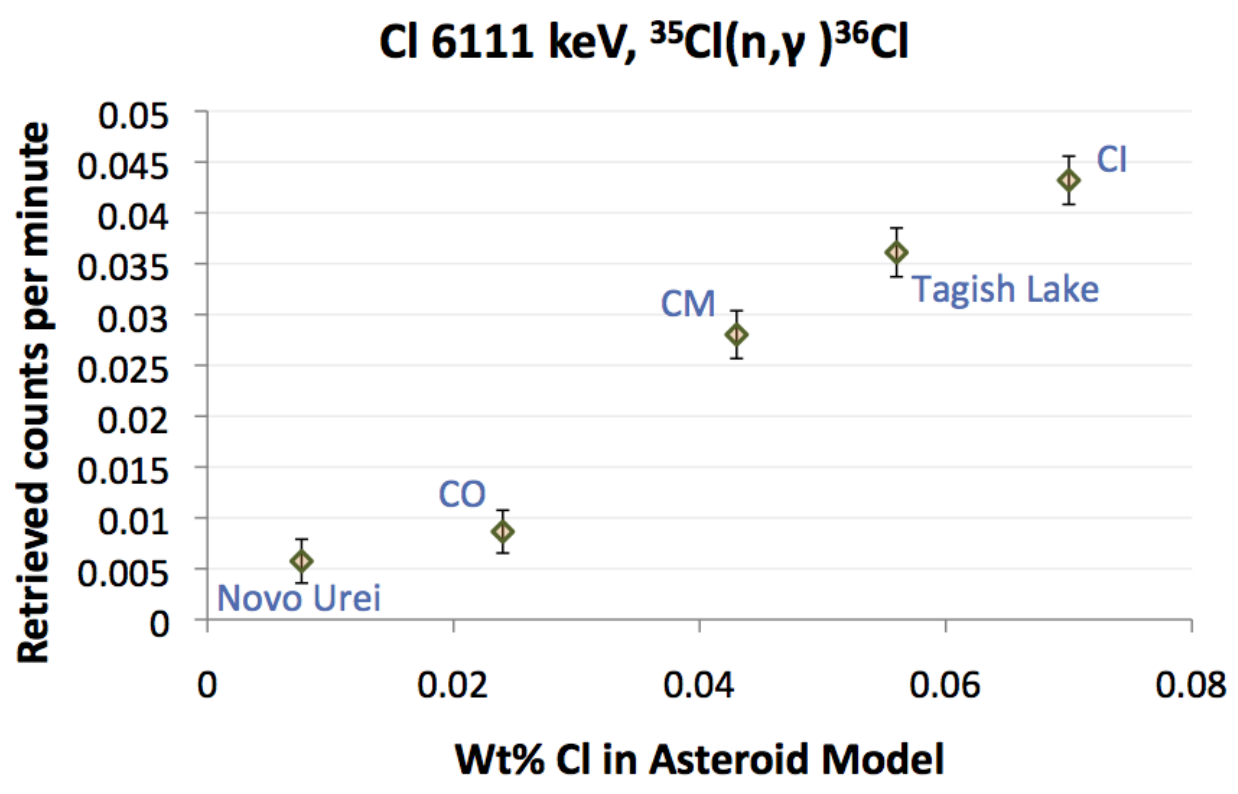

**Figure 9. Retrieved count rate in the 6111 keV line of chlorine vs. model asteroid composition.** Error bars are derived from peak fitting and represent photon statistical uncertainties ($\sigma$) after 30 days of integration.

#### 3.1.5. *Carbon, Oxygen, and the 4438 keV Region*

The strong 4438 keV line produced by $^{12}$C (n,n′$\gamma$) and $^{16}$O(n,n′$\alpha\gamma$)$^{12}$C is Doppler broadened by the nuclear recoil of its production. Measurement of carbon abundance based on this line is challenging but achievable (*Peplowski et al.* 2015). Overlapping lines from iron, silicon, and aluminum sit on the broad wings of the C+O peak (Fig. 10). The carbon count rate in the 4438 keV line may be determined by subtracting the oxygen contribution, which can be calculated based on the count rate in the 6129 keV line of $^{16}$O (n,n′$\gamma$) (Table 6).

**Table 6**. Carbon Sensitivity Estimation and Error Budget for a CI Chondritic Asteroid Composition

| | Counts per Minute ($\pm 3\sigma$ after 135 days) |
|---|---|
| Total 4438 keV count rate, retrieved from model spectrum = $C_{Total}$(4438): | 0.431 $\pm$ 0.016 |
| Spacecraft 4438 keV count rate, retrieved from model background[a] = $C_{Spacecraft}$(4438): | 0.038 $\pm$ 0.011 |
| Asteroid-originating 4438 keV count rate = $C_{Total}$ - $C_{Spacecraft}$ $\equiv$ $C_{Asteroid}$(4438): | 0.393 $\pm$ 0.020 |
| | |
| Asteroid-originating 6129 keV $^{16}$O (n,n′$\gamma$) count rate, retrieved = $C_O$(6129) | 0.266 $\pm$ 0.004 |
| MCNPX oxygen (single-element spectrum) line ratio, $C_O$(4438)/$C_O$(6129) = 0.95[b] | |
| Calculated $O_{4438}$ count rate = ($C_O$(6129))*($C_O$(4438)/$C_O$(6129)) = $C_O$(4438 keV): | 0.253 $\pm$ 0.004 |
| Asteroid Carbon (4438 keV) = $C_{Asteroid}$(4438) - $C_O$(4438 keV): | 0.140 $\pm$ 0.020 |

*Table 6 continued*

Table 6 *(continued)*

| Counts per Minute ($\pm$ 3$\sigma$ after 135 days) |
|---|

[a] In reality only the GCR component of the spacecraft background can be measured by direct spectral fitting. For one technique for determining the neutron/spacecraft component of the background see (*Peplowski et al.* 2015). Note also that the neutron/spacecraft component is much smaller in a one-radius asteroid orbit than it was for MESSENGER or Mars Odyssey because of the lower solid angle ($\Omega$) occupied by the asteroid (Figure 3).

[b] This ratio is significantly higher than that determined by *Peplowski et al.* (2015) via integration over the neutron cross sections, even accounting for the differences in detector efficiency between the two instruments. Thus, MCNPX is overestimating the background due to oxygen in this line and its associated uncertainties. However, as is evident from the table, this is not the dominant component of the error budget.

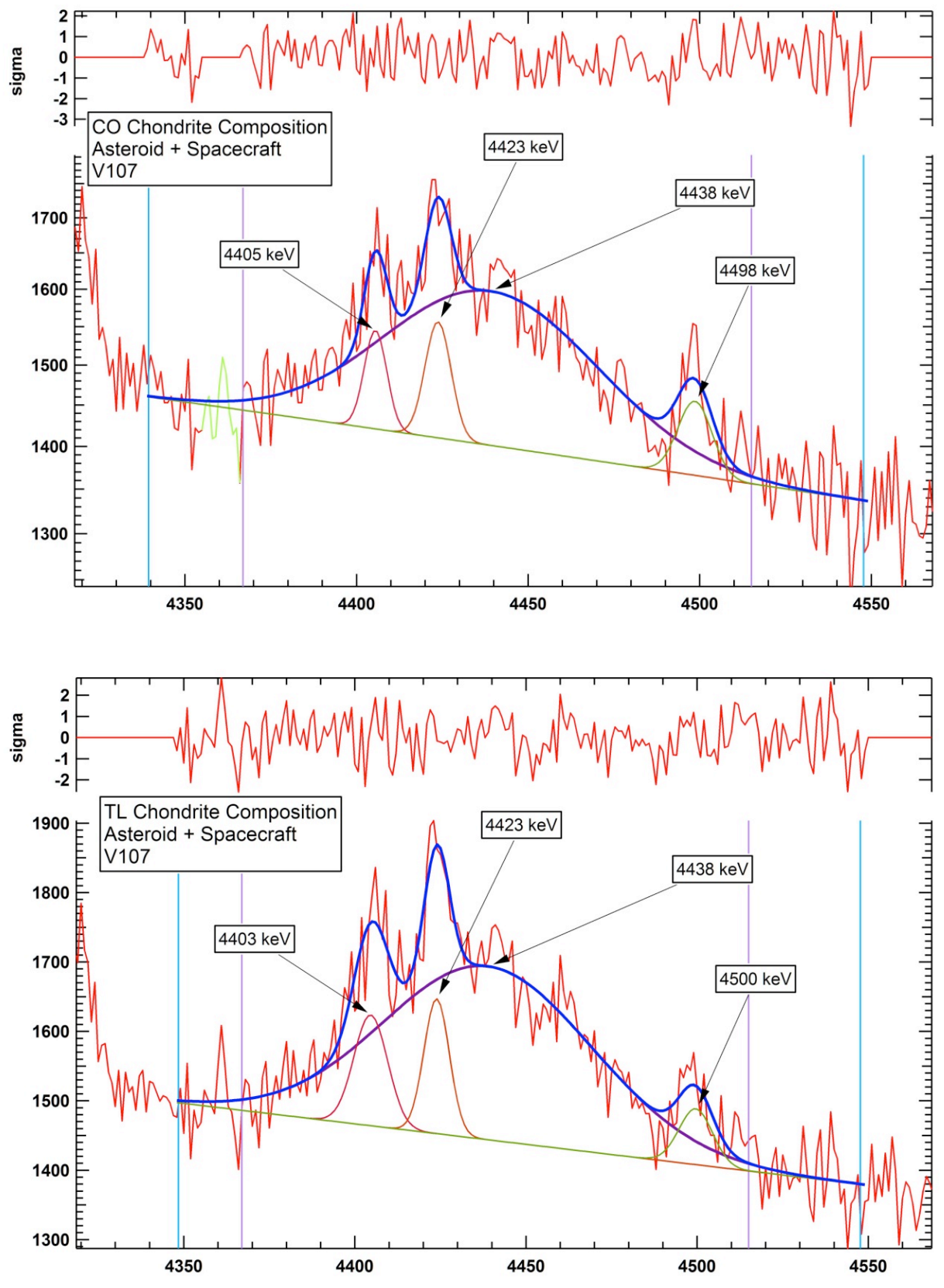

**Figure 10**. Fitting of the gamma-ray lines in the complex $^{12}$C (n,n′$\gamma$) and $^{16}$O(n,n′$\alpha\gamma$)$^{12}$C region (4438 keV) of the CO chondrite and Tagish Lake model spectra. The large Doppler-broadened peak is the combined $^{12}$C + $^{16}$O signal. Peak identifications are as follows: 4403/4405 keV, $^{56}$Fe(n,$\gamma$) (nominal energy 4406.1 keV); 4423 keV, escape peak of $^{28}$Si(n,$\gamma$) at 4934 keV; 4438 keV, $^{12}$C (n,n′$\gamma$) and $^{16}$O(n,n′$\alpha\gamma$)$^{12}$C; 4498/4500 keV, multiple sources including $^{28}$Si(n,n′$\gamma$) and $^{27}$Al(p,$\gamma$)$^{28}$Si (nominal energy 4496.8 keV). See *Evans et al.* (2006); *Reedy* (1978); *Antilla et al.* (1977).

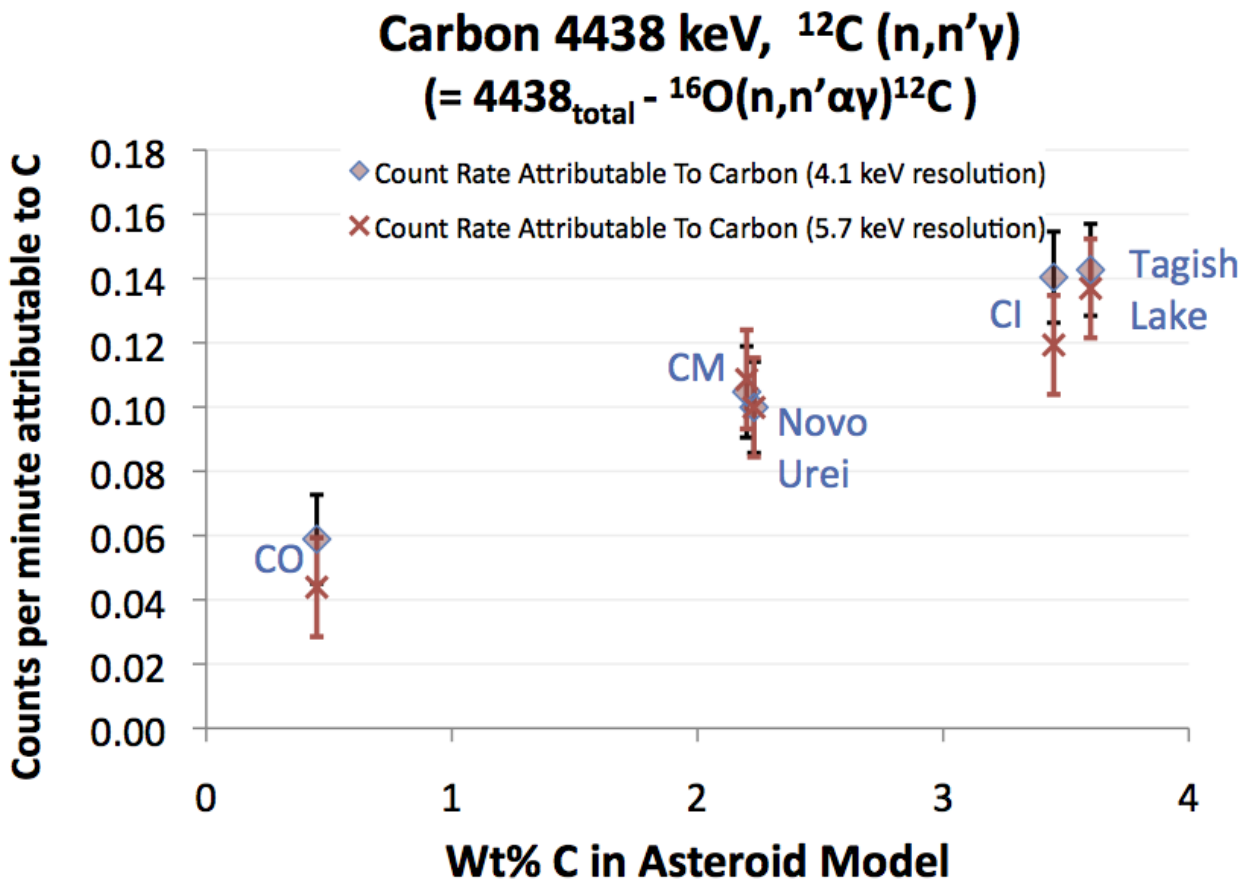

**Figure 11**. Retrieved count rates attributable to carbon in the 4438 keV line vs. model asteroid composition. Error bars are derived from peak fitting and represent photon statistical errors ($\sigma$) after 30 days of integration. The C contribution to the 4438 keV peak is higher in count rate than the 5420 keV S peak (Fig. 8) but the C uncertainties are much higher both because of the Doppler broadening and because each of the many additional lines that must be analyzed contributes to the error budget.

We find that an integration time of 135 days in a Dawn/LAMO-like orbit is sufficient to characterize the carbon abundances in carbon-rich meteorite compositions and to set meaningful upper limits on the abundances in carbon-poor compositions (Figure 11, Table 2).

### 3.2. *Validation: comparison of model results with Mars Odyssey GRS flight data*

We compared our model line count rates against the Mars Odyssey count rates (*Evans et al.* 2006) for the two models closest to Mars Odyssey in hydrogen content, as the hydrogen abundance has a greater effect than that of any other element on the thermal neutron population.

Our model asteroid orbiter has a less favorable geometry than the Odyssey experiment ($\Omega$ = 0.84 steradians, vs. 3.47 steradians of solid angle at Mars), but benefits from the absence of the Martian atmosphere, which attenuates both incoming cosmic rays and outgoing $\gamma$-rays. After accounting for the atmospheric effects (*Masarik and Reedy* 1996) and the differences in geometry and abundances of $\gamma$-ray source nuclides, our CO chondrite model count rates (Table 7) are consistent with the Mars Odyssey midlatitude flight count rates to within ±45% over a wide range of $\gamma$-ray energies. This is as expected given the significant differences in the neutron population caused by the much lower abundances of Cl and much higher abundances of Fe in the chondrite models as compared to Mars.

Table 8 gives a similar comparison for the next model up in hydrogen concentration, the CM chondrite model. The super-Martian hydrogen abundance drives up the thermal neutron population in the Monte-Carlo model, thus elevating all the capture lines except that of hydrogen itself relative to the scaled Mars Odyssey count rates. Meanwhile, the inelastic lines are reduced in strength compared to the CO model, most likely because of the lower Fe concentration in the CMs.

Note that count rates in the 4438 keV line in the Odyssey flight data are always dominated by emission from the Martian $CO_2$ atmosphere and thus not informative about the surface composition.

**Table 7**. Mars Odyssey GRS Data vs. CO Chondrite Model $\gamma$-Ray Line Count Rates, in Counts per Minute ("CPM")

| | $\gamma$-Ray Energy (keV) | CO Model CPM[a] | Mars ice-free model wt%[b] | CO model wt% | Ody. GRS count rate (CPM)[c] | Geometric Factor[d] | Mars Atmos. Factor[e] | Ody. CPM Scaled to CO[f] | Ratio: CO model to scaled Ody. GRS data |
|---|---|---|---|---|---|---|---|---|---|
| | | | | Neutron Capture $\gamma$-Rays | | | | | |
| H | 2224 | 0.099 | 0.32 | 0.07 | 1.077 | 0.242 | 2.69 | 0.154 | 0.64 |
| Si | 3539 | 0.072 | 21.33 | 15.90 | 0.228 | 0.242 | 2.15 | 0.089 | 0.81 |
| Si | 4423 | 0.028 | 21.33 | 15.90 | 0.121 | 0.242 | 1.84 | 0.040 | 0.70 |
| Si | 4934 | 0.067 | 21.33 | 15.90 | 0.223 | 0.242 | 1.84 | 0.074 | 0.91 |
| S | 5420 | 0.020 | 2.20 | 2.00 | 0.038 | 0.242 | 1.76 | 0.015 | 1.36 |
| Cl | 6111 | 0.007 | 0.54 | 0.024 | 0.323 | 0.242 | 1.66 | 0.006 | 1.21 |
| Fe | 7631 | 0.233 | 14.60 | 24.80 | 0.336 | 0.242 | 1.49 | 0.206 | 1.13 |
| Fe | 7646 | 0.200 | 14.60 | 24.80 | 0.306 | 0.242 | 1.49 | 0.188 | 1.07 |
| | | | | Inelastic Scatter $\gamma$-Rays | | | | | |
| Si | 1779 | 1.358 | 21.33 | 15.90 | 1.773 | 0.242 | 2.97 | 0.952 | 1.43 |
| O | 6127 | 0.264 | 45.30 | 37.00 | 0.952 | 0.242 | 1.66 | 0.312 | 0.85 |
| | | | | | | | | Median: | 1.07 |
| | | | | | | | | $\sigma$: | 0.27 |

*Table 7 continued*

Table 7 *(continued)*

| γ-Ray Energy (keV) | CO Model CPM[a] | Mars ice-free model wt%[b] | CO model wt% | Ody. GRS count rate (CPM)[c] | Geometric Factor[d] | Mars Atmos. Factor[e] | Ody. CPM Scaled to CO[f] | Ratio: CO model to scaled Ody. GRS data |
|---|---|---|---|---|---|---|---|---|

[a] CO chondrite-based asteroid model line count rates from Table 5

[b] Average composition of ice-free latitudes from *Boynton et al.* (2007)

[c] Odyssey midlatitude count rates from *Evans et al.* (2006)

[d] Ratio of asteroid solid angle Ω=0.84 sr to Mars Odyssey solid angle Ω=3.47 sr

[e] Includes cosmic-ray production and atmospheric absorption effects (*Masarik and Reedy* 1996)

[f] Scaled for atmosphere, composition, and solid angle.

**Table 8**. Mars Odyssey GRS Data vs. CM Chondrite Model γ-Ray Line Count Rates, in Counts per Minute ("CPM")

| | γ-Ray Energy (keV) | CM Model CPM[a] | Mars ice-free model wt%[b] | CM model wt% | Ody. GRS count rate (CPM)[c] | Geometric Factor[d] | Mars Atmos. Factor[e] | Ody. CPM Scaled to CM[f] | Ratio: CM model to scaled Ody. GRS data |
|---|---|---|---|---|---|---|---|---|---|
| | | | | | Neutron Capture γ-Rays | | | | |
| H | 2224 | 1.208 | 0.32 | 1.40 | 1.077 | 0.242 | 2.69 | 3.071 | 0.39 |
| Si | 3539 | 0.114 | 21.33 | 12.70 | 0.228 | 0.242 | 2.15 | 0.071 | 1.60 |
| Si | 4423 | 0.041 | 21.33 | 12.70 | 0.121 | 0.242 | 1.84 | 0.032 | 1.28 |
| Si | 4934 | 0.096 | 21.33 | 12.70 | 0.223 | 0.242 | 1.84 | 0.059 | 1.63 |
| S | 5420 | 0.048 | 2.20 | 2.70 | 0.038 | 0.242 | 1.76 | 0.020 | 2.42 |
| Cl | 6111 | 0.024 | 0.54 | 0.043 | 0.323 | 0.242 | 1.66 | 0.010 | 2.32 |
| Fe | 7631 | 0.392 | 14.60 | 21.30 | 0.336 | 0.242 | 1.49 | 0.177 | 2.21 |
| Fe | 7646 | 0.341 | 14.60 | 21.30 | 0.306 | 0.242 | 1.49 | 0.161 | 2.11 |
| | | | | | Inelastic Scatter γ-Rays | | | | |
| Si | 1779.90 | 0.846 | 21.33 | 12.70 | 1.773 | 0.242 | 2.97 | 0.760 | 1.11 |
| O | 6127.00 | 0.258 | 45.30 | 43.20 | 0.952 | 0.242 | 1.66 | 0.364 | 0.71 |
| | | | | | | | | Median: | 1.63 |
| | | | | | | | | σ: | 0.70 |

[a] CM chondrite-based asteroid model line count rates from Table 5

[b] Average composition of ice-free latitudes from *Boynton et al.* (2007)

[c] Odyssey midlatitude count rates from *Evans et al.* (2006)

[d] Ratio of asteroid solid angle Ω=0.84 sr to Mars Odyssey solid angle Ω=3.47 sr

[e] Includes cosmic-ray production and atmospheric absorption effects (*Masarik and Reedy* 1996)

[f] Scaled for atmosphere, composition, and solid angle.

## 3.3. *Carbon Sensitivity in the MESSENGER GRS flight experiment*

Like the Mars Odyssey GRS, the MESSENGER GRS experiment (*Goldsten et al.* 2007; *Evans et al.* 2012) used

a high-purity germanium detector with the high energy resolution required to resolve lines in the complex 4438 and 2223 keV regions where carbon and hydrogen can be measured (Table 4, "Flight Detector Resolution") as demonstrated by *Peplowski et al.* (2015). Unfortunately, *Peplowski et al.* found that the carbon sensitivity of the MESSENGER experiment was limited to $\pm 2.7$ wt% ($3\sigma$) primarily by the background signal from the large carboniferous plastic anticoincidence filter surrounding the detector (0.14 counts per minute in the 4438 keV line) and thus did not obtain a confident detection of carbon in the GRS spectrum of Mercury. Without the spacecraft background, they estimate that carbon would have been detectable at $\approx 0.5$ wt%.

The large size (6.7 cm diameter x 6.7 cm height right circular cylinder) of the Mars Odyssey GRS gave it greater $\gamma$-ray stopping power than the smaller (5 cm x 5 cm) MESSENGER GRS in the 4438 and 6129 keV lines needed for the carbon measurement.

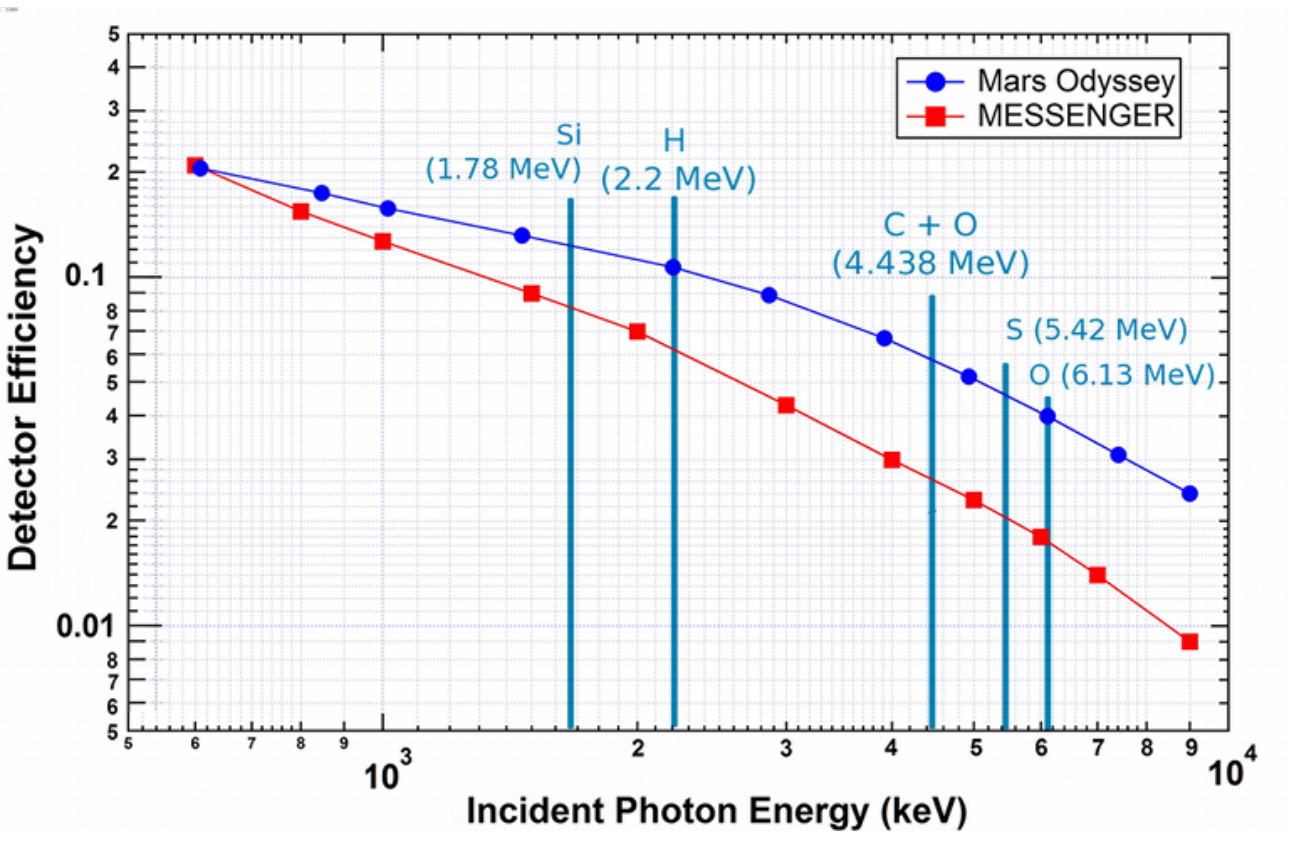

**Figure 12.** Detection efficiencies of two successful heritage HPGe (high-resolution) gamma-ray instruments: the 6.7 x 6.7 cm cylindrical Mars Odyssey Gamma-Ray Spectrometer and the 5-cm MESSENGER Gamma-Ray Spectrometer (*Goldsten et al.* 2007). The larger size of the Odyssey GRS enables high sensitivity to high-energy gamma-ray lines, including the 4.4 MeV line of carbon and the 6.13 MeV line of oxygen. Both of these lines are needed to determine carbon abundance.

## 4. CONCLUSIONS

A major objective of any small-body orbiter mission will be to illuminate the relationship between the materials in the target object and the meteorites that have been delivered to Earth. Currently, the parent bodies of the various carbonaceous chondrite subclasses and their relationships to one another remain unknown. We find that global measurements of sulfur, carbon, hydrogen, and other elements with sufficient accuracy to establish undifferentiated composition and carbonaceous subclass can be achieved with a heritage gamma-ray instrument, and that the integration time required for a successful experiment at a feasible orbital altitude are consistent with the operational time frames of prior NASA small-body missions.

## 5. ACKNOWLEDGEMENTS

## 6. SUPPLEMENTAL MATERIAL

Not every element in Table S1 produces a $\gamma$-ray line measurable from orbit, but every element does influence the neutron energy distribution—sometimes very substantially, as with the rare-earth elements Sm and Gd. For this reason, we attempted to provide accurate inputs to MCNPX for several minor and trace elements.

**Table S1**. Model Asteroid Compositions As Input to MCNPX (wt%)

| Element | CI | Tagish Lake | CM | CO | Novo Urei |
|---|---|---|---|---|---|
| H | 2.02 | 1.5 | 1.4 | 0.07 | 0 |
| C | 3.45 | 3.6 | 2.2 | 0.45 | 2.23 |
| N | 0.2 | 0.1218 | 0.0967 | 0.009 | 0 |
| O-16[a] | 46.2024 | 45.12281 | 43.28351336 | 36.608 | 38.685 |
| O-17[a] | 0.017598 | 0.017187 | 0.016486644 | 0.013944 | 0.015 |
| Na | 0.5 | 0.445 | 0.39 | 0.41 | 0.13 |
| Mg | 9.7 | 10.8 | 11.5 | 14.5 | 22.23 |
| Al | 0.865 | 0.99 | 1.13 | 1.43 | 0.26 |
| Si-28 | 9.813272 | 10.51422 | 11.71320999 | 14.665 | 17.127 |
| Si-29 | 0.496888 | 0.53238 | 0.593090009 | 0.74253 | 0.867 |
| Si-30 | 0.32984 | 0.3534 | 0.393699997 | 0.4929 | 0.576 |
| P | 0.095 | 0.0927 | 0.103 | 0.104 | 0.04 |
| S | 5.41 | 3.8 | 2.7 | 2 | 0.58 |
| Cl | 0.07 | 0.056 | 0.043 | 0.024 | 0.00765 |
| K | 0.055 | 0.065 | 0.037 | 0.0345 | 0.03 |
| Ca | 0.926 | 0.99 | 1.29 | 1.58 | 0.57 |
| Sc | 0 | 0 | 0 | 0 | 0.000909 |
| Ti | 0.044 | 0.052 | 0.055 | 0.078 | 0.08 |
| Cr-50 | 0.011514 | 0.01234 | 0.01325225 | 0.015425 | 0.02 |
| Cr-52 | 0.222041 | 0.237961 | 0.255556452 | 0.29745 | 0.394 |
| Cr-53 | 0.025178 | 0.026983 | 0.028978048 | 0.033729 | 0.045 |
| Cr-54 | 0.006267 | 0.006717 | 0.00721325 | 0.0083957 | 0.011 |
| Mn-55 | 0.194 | 0.149 | 0.165 | 0.165 | 0.31 |
| Fe-54 | 1.055494 | 1.119288 | 1.235276395 | 1.4383 | 0.907 |
| Fe-56 | 16.69137 | 17.70019 | 19.53440666 | 22.744 | 14.344 |
| Fe-57 | 0.40218 | 0.426487 | 0.470682915 | 0.54803 | 0.346 |
| Fe-58 | 0.050955 | 0.054035 | 0.059634033 | 0.069433 | 0.044 |
| Co | 0.0505 | 0.0517 | 0.056 | 0.0688 | 0.05 |
| Ni | 1.1 | 1.16 | 1.23 | 1.4 | 0.12 |
| Cu | 0 | 0 | 0 | 0 | 0.022 |
| Zn | 0 | 0 | 0 | 0 | 0.134 |
| Ga | 0 | 0 | 0 | 0 | 0.00028 |
| Ge | 0 | 0 | 0 | 0 | 0.0014 |
| Se | 0 | 0 | 0 | 0 | 0 |

*Table S1 continued*

Table S1 *(continued)*

| Element | CI | Tagish Lake | CM | CO | Novo Urei |
|---|---|---|---|---|---|
| Sm | 0.000015 | 0.00002 | 0.0000204 | 0.000024 | 0.00000107 |
| Gd | 0.00002 | 0.000024 | 0.000029 | 0.0000337 | 0.00000261 |
| Th-232 | 0.000003 | 0.000004 | 0.0000041 | 0.000006 | 0.000000145 |
| U-234 | 0.000001 | 0.000001 | 0.0000012 | 0.0000013 | 0.0000001 |
| Total[a]: | 100.00 | 100.00 | 100.00 | 100.00 | 100.18 |

[a]Total oxygen values were derived by subtracting the sum of the remaining elements from 100%.

NOTE— Meteorite models were based on published analyses or averages as follows: CI, Tagish Lake, CM— Table 2 of *Brown et al.* (2000); CO—Subclass average from *Wasson and Kallemeyn* (1988); Novo Urei— *Wiik* (1972) (major elements); REE and trace elements from *Goodrich et al.* (1991); *Spitz and Boynton* (1991). Elements not listed were treated by MCNPX as having zero abundance.

**Table S2**. Spurious Gamma-Ray Lines Removed from the MCNPX Output

| Primary Science $\gamma$-ray Affected | GCR Spacecraft Background Lines Removed (keV) | Neutron Spacecraft Background Lines Removed (keV) |
|---|---|---|
| 1779 keV Si line | 1774, 1794 | 1773, 1774, 1775, 1776, 1793, 1794, 1795 |
| 1942 keV Ca line | 1941, 1942, 1943, 1948, 1949, 1951, 1952, 1960, 1967, 1971 | 1941, 1945, 1949, 1951, 1959, 1962, 1963, 1965, 1967 |
| 2223 keV H line | 2229, 2231, 2234, 2237, 2245, 2247 | 2245 |
| 4438 keV C+O line | 4410, 4439, 4446, 4511 | 4377, 4378, 4411, 4439, 4446, 4511, 4512 |
| 6129 keV O line | 6094, 6130, 6131 | 6094, 6095, 6130, 6131 |